\documentclass[a4paper,11pt]{article}
\pdfoutput=1 

\usepackage{jheppub} 

\usepackage[T1]{fontenc} 

\usepackage{pdfsync}
\usepackage{mathptmx}
\usepackage[utf8]{inputenc}
\usepackage[T1]{fontenc}
\usepackage{slashed}
\usepackage{times}
\usepackage{amssymb,amsfonts,amsmath,amsthm}
\usepackage{dsfont,bbm}
\usepackage{dcolumn}
\usepackage{epsf}
\usepackage{graphicx}
\usepackage[caption=false]{subfig}
\usepackage{dsfont}
\usepackage{bm}
\usepackage{eucal}
\usepackage{slashed}
\usepackage[active]{srcltx}
\usepackage[usenames]{color}

\title{\boldmath Vacuum Integration: UV- and IR-divergencies}

\author[a]{I.~V.~Anikin}

\affiliation[a]{Bogoliubov Laboratory of Theoretical Physics JINR, 141980 Dubna, Russia}

\emailAdd{anikin@theor.jinr.ru}

\abstract{In this note we present the important details regarding the massless vacuum integrations
which are not outlined in the literature. 
In particular, it has been shown how the delta-function represents either UV-regime or IR-regime.
In the case of vacuum integration, we advocate the use of sequential approach to the
singular generated functions (distributions). 
The sequential approach is extremely useful for many practical applications, 
in particular, in the effective potential method.}

\begin{document}
\maketitle
\flushbottom

\section{Introduction}
\label{Intro}

In different QFT-models, at the classical level,
the effects of  spontaneous symmetry breaking are very important in the context of
the geometrical analysis of the Goldstone theorem.
In this connection, the study of a vacuum state as
the potential minimum plays an significant
role  \cite{Vasilev:2004yr}.
Meanwhile, the quantum corrections, that tend usually to
distort the geometrical picture, computed within the effective potential (EP) methods allow to
return again to the classical geometrical analysis of the models with spontaneous symmetry breaking.
In the standard EP-approaches, the quantum corrections are given by the
the vacuum integrations with the massive propagators.
However, the special interest is related to the vacuum integrations with the massless propagators.
It is mostly dictated by the use of conformal symmetry (see for example \cite{Anikin:2020dlh, Anikin:2023wkk, Anikin:2023ogb}).

On the other hand, working with the vacuum massless integrations, it demands some careful considerations.
Indeed, the general dimensional analysis suggests that all vacuum integrations with the massless propagators
lead to zero \cite{Grozin:2005yg, Grozin:2007zz}. It is true except a particular case of dimensionless integrand
where the ultraviolet (UV), or infrared (IR), momentum region is only under consideration.
In this case, the arguments of dimensional analysis cannot be applied.

In \cite{Gorishnii:1984te}, it has been shown that the vacuum integration of dimensionless and massless integrand
is proportional to $\delta(n-D/2)$ where the space dimension is defined as $D=d-2\varepsilon$
($d=2, 4, 6\,\, \text{etc.}$) and $n$ implies the propagator index.
The delta-function as a singular generated function (distribution) is a well-defined linear functional on the
suitable finite function space. In the case of dimensional regularization, this space should be realized with the
integration measure as $d\varepsilon \, \varphi(\varepsilon)$ where $\varphi$ has a localized support.
However, it is not always convenient, even possible, to deal with the measure as $d\varepsilon$
\cite{Anikin:2023wkk, Anikin:2023ogb}. Moreover, 
owing to the symmetry properties, the delta-function is usually hiding the information on
the UV(or IR)-divergency.

Following Gorishni-Isaev's method \cite{Gorishnii:1984te}, 
we present all necessary details on the vacuum integration where the delta-function
has been treated in the frame of the sequential approach \cite{Antosik:1973,Gelfand:1964}. 
We also demonstrate how the delta-function represents  the UV(IR)-regimes.

\section{$\Delta_F(0)$-singularity}
\label{DeltaF-1}

Let us consider the simplest case of scalar massless propagator $\Delta_F(0)$ giving the tad-pole diagram. 
Using the Fourier transform, the propagator $\Delta_F(0)$ can be write as
\footnote{For the sake of shortness, here and in what follows the momentum loop normalization is hidden in $(d^D k)$.
Moreover, the Euclidian measure of momentum integrations has been implies.}
\begin{eqnarray}
\label{V-Int-1}
&&\Delta_F(0) = \int \frac{(d^D k)}{k^2}=
\int (d^D k) \Big\{ C^{-1}(D,1) \int d^Dz\, \frac{e^{-i k z}}{\big(z^2\big)^{D/2-1}} \Big\}
\nonumber\\
&&=C^{-1}(D,1) \int d^Dz \, \,\frac{\delta(z)}{\big(z^2\big)^{D/2-1}}\equiv
\Gamma(D/2-1) \int (d^Dz) \, \,\frac{\delta(z)}{\big(z^2\big)^{D/2-1}},
\end{eqnarray}
where the integration measure $(d^D z)$ absorbs the normalization constant 
$i(-\pi)^{D/2}$ arising from 
\begin{eqnarray}
\label{C}
C^{-1}(D,n)=i(-\pi)^{D/2}\frac{\Gamma(D/2-n)}{\Gamma(n)}.
\end{eqnarray}
If we assume that $D/2-1=0$,
then the propagator in Eqn.~(\ref{V-Int-1}) takes a form of
\begin{eqnarray}
\label{V-Int-2}
\Delta_F(0) = \Gamma(0) \int (d^Dz) \, \delta(z) \Rightarrow \Gamma(0),
\end{eqnarray}
where, as well-known, the singularity of $\Gamma$-function can be presented as
\begin{eqnarray}
\label{V-Int-3}
 \Gamma(0) = \lim_{\epsilon\to 0} \Gamma(\epsilon)=  \lim_{\epsilon\to 0} \Big\{ \frac{1}{\epsilon} + ....\Big\}.
\end{eqnarray}
It is worth to notice that
the condition given by $D/2-1=0$ should 
be applied before the integration over $(d^D k)$ in Eqn.~(\ref{V-Int-1}) in order to avoid the uncertainty, see also Sec.~\ref{DeltaF-2}.

On the other hand, according to \cite{Gorishnii:1984te}, the vacuum integration method applied to
the Feynman propagator results in the delta-function. Let us remind a key moment of Gorishni-Isaev's method.
Using the spherical system (in the momentum Euclidian space),  $\Delta_F(0)$ can be represented as
\begin{eqnarray}
\label{V-Int-1-w1}
&&\Delta_F(0) = \int \frac{(d^D k)}{k^2}=
\frac{1}{2} \int d\Omega \int_{0}^\infty d\beta \, \beta^{D/2-2},
\end{eqnarray} 
where $d\Omega$ gives the finite angle measure of integration.
The replacement $\beta = e^y$ leads to the following expression
\begin{eqnarray}
\label{V-Int-1-w2}
&&\Delta_F(0) = 
\frac{1}{2} \int d\Omega \int_{-\infty}^\infty (dy) \, e^{iy \big[(-i)(D/2-1)\big]}=
\frac{1}{2\, | i |}\delta\big( D/2 -1 \big)\, \int d\Omega
\end{eqnarray}  
or, restoring all coefficients, it reads
\begin{eqnarray}
\label{V-Int-3-2}
\Delta_F(0)= - \, 2i\, \pi^{1+D/2} \,\delta(1-D/2) \Big|_{D=2} = - \, 2i\, \pi^{2} \,\delta(0).
\end{eqnarray}
So, for the case of $D=2$, the matching of Eqns.~(\ref{V-Int-2}) and (\ref{V-Int-3-2}) gives the following
representation
\begin{eqnarray}
\label{V-Int-4}
(-i) \,  \Delta_F(0)=\Gamma(0) = - \, 2\, \pi^{2} \,\delta(0).
\end{eqnarray}
With this, we may conclude that $\delta(0)$-singularity can be treated as the singularity of $\Gamma(0)$, see Eqn.~(\ref{V-Int-3}).
The same inference has been reached by the different method, see \cite{Anikin:2020dlh}.
Notice that the physical (UV or IR) nature of the mentioned singularity has been somewhat hidden.

In the dimensional regularization, the UV- and IR-divergencies are associated with the small positive ($\varepsilon >0$)
and negative ($\varepsilon < 0$) regularized parameter $\varepsilon$, respectively.
In this connection, using the $\alpha$-parametrization, we rewrite Eqn.~(\ref{V-Int-1}) as
\begin{eqnarray}
\label{V-Int-1-2}
&&\Delta_F(0) = \int \frac{(d^D k)}{k^2}=
\Gamma(D/2-1) \int (d^Dz) \, \,\frac{\delta(z)}{\big(z^2\big)^{D/2-1}}
\nonumber\\
&&=
\int (d^Dz) \, \,\delta(z) \, \Big\{
\int_{0}^{\infty} d\alpha\, \alpha^{D/2-2} \, e^{-\alpha z^2}
\Big\}=\int_{0}^{\infty} (d\alpha)\, \alpha^{D/2-2}.
\end{eqnarray}
Hence, one gets (modulo the normalization factor which is now irrelevant)
\begin{eqnarray}
\label{V-Int-1-3}
\Delta_F(0) = \int \frac{(d^D k)}{k^2}=
\int_{0}^{\infty} (d\alpha)\, \alpha^{D/2-2}\Rightarrow
\frac{1}{D/2-1} \Big\{
\lim_{\alpha\to\infty} \alpha^{D/2-1} - \lim_{\alpha\to 0 } \alpha^{D/2-1}
\Big\}.
\end{eqnarray}
From Eqn.~(\ref{V-Int-1-3}), one can see that the first term corresponds to the UV-divergency, while the second term --
to the IR-divergency. That is, we have
\begin{eqnarray}
\label{UV-IR-1}
&&\lim_{\alpha\to\infty} \alpha^{D/2-1} = [\infty]_{\text{UV}} \quad \text{if} \,\,\, D >2,
\nonumber\\
&&\lim_{\alpha\to 0} \,\,\alpha^{D/2-1} = \,\,[\infty]_{\text{IR}} \quad \text{if} \,\,\, D < 2.
\end{eqnarray}
In other words, if the dimensional parameter $\epsilon$ in $D= d - 2\epsilon$ is small one, $| \epsilon | < 1$,
 and it varies from the negative to positive variables,
we have the following representation for $\Delta_F(0)$
 \begin{eqnarray}
\label{V-Int-1-3}
\Delta_F(0) \Big|_{d=2}
&&\Rightarrow
\frac{1}{D/2-1} \Big\{
\Theta(D>2 \,| \, \epsilon < 0)
\lim_{\alpha\to\infty} \alpha^{D/2-1} -
\Theta(D<2\,| \, \epsilon > 0)
\lim_{\alpha\to 0 } \alpha^{D/2-1}
\Big\} = 0
\nonumber\\
&&
\Rightarrow \delta\big( 1- D/2 \big) \Big|_{D\not= 2}=0,
\end{eqnarray}
where $\epsilon$ should be considered as an external independent parameter.
From Eqns.~(\ref{UV-IR-1}) and (\ref{V-Int-1-3}), in the dimensional regularization, one can see that the 
positive small $\epsilon$ is regularizing the UV-divergency but not IR-divergency.
Thus,  every of the methods gives the same final conclusion.

To conclude this section, we remind the other useful representation given by 
\begin{eqnarray}
\label{l-c-delta}
\Delta_F(0) = \lim_{z^2\to 0} \Delta_F(z^2) =
\lim_{z^2\to 0} \frac{1}{4\pi} \delta_+(z^2)= \delta(0), \quad z\in \mathbb{E}^4
\end{eqnarray}
which is in agreement with Eqns.~(\ref{V-Int-3-2}) and (\ref{V-Int-4}).

\section{Vacuum integration as a limit of non-vacuum integration}
\label{DeltaF-2}

We now address to the relation between vacuum and non-vacuum integrations.
In the dimensional regularization procedure, we begin with the consideration of
two-point 1PI massless Green function given by
\begin{eqnarray}
\label{V-nV-1}
\mathcal{I}(p^2)= \int \frac{(d^D k)}{k^2(k^2+p^2)}=
(c.c.)\,(p^2)^{D/2-2} \, G(1,1),
\end{eqnarray}
where $(c.c)$ implies the coefficient constant and
\begin{eqnarray}
\label{G11}
G(1,1)=\frac{\Gamma(-D/2+2) \Gamma^2(D/2-1)}{\Gamma(D-2)}.
\end{eqnarray}
Using $D=4-2\epsilon$, we get
\begin{eqnarray}
\label{V-nV-1-2}
\mathcal{I}(p^2)= \int \frac{(d^D k)}{k^2(k^2+p^2)}=(c.c.)\,(p^2)^{-\epsilon} \,
\frac{\Gamma(\epsilon) \Gamma^2(1-\epsilon)}{\Gamma(2-2\epsilon)}.
\end{eqnarray}
In Eqns.~(\ref{V-nV-1}) and (\ref{V-nV-1-2}), the scale dependence of $\mu^2$ is hidden as irrelevant one.

The vacuum integration can be obtained from Eqn.~(\ref{V-nV-1}) with the help of the corresponding limit as
\begin{eqnarray}
\label{V-nV-2}
&&\mathcal{V}_2\equiv \int \frac{(d^D k)}{(k^2)^2}=
\lim_{p^2\to 0}
\mathcal{I}(p^2).
\end{eqnarray}
There are, however, some subtleties of this limit which are now under our considerations.
Indeed, having used the $\alpha$-representation, let us calculate the integral of Eqn.~(\ref{V-nV-1}).
We have the following
\begin{eqnarray}
\label{V-nV-3}
\mathcal{I}(p^2)= (c.c.)
\int_0^\infty d\alpha d\beta \frac{e^{-p^2 \frac{\alpha \beta }{\alpha + \beta}}}{[\alpha+\beta]^{D/2}}
=(c.c.) \int_0^\infty \lambda \lambda^{1-D/2}
\int_0^1 dx e^{-p^2\lambda x \bar x},
\end{eqnarray}
where
\begin{eqnarray}
\label{Al-Rep-1}
\alpha =\lambda x_1, \quad \beta=  \lambda x_2, \quad \lambda\in [0, \, \infty].
\end{eqnarray}
The next stage of calculations is to make a replacement as
\begin{eqnarray}
\label{Rep-1}
\tilde\lambda= p^2 \lambda x \bar x, \quad d\tilde\lambda = p^2 x \bar x d\lambda
\end{eqnarray}
in the exponential function. This replacement simplifies the integrals and it leads to the corresponding combination
of $\Gamma$-functions denoted as $G(1,1)$ \cite{Grozin:2005yg, Grozin:2007zz}. Ultimately, we reproduce the result
presented by Eqns.~(\ref{V-nV-1}) and (\ref{V-nV-1-2}).

Now, the first mathematical subtlety is that if we suppose the limits $p^2\to 0$ and $\epsilon\to 0$
are consequent ones, not simultaneous, it is clear that these limits are not commutative operations, {\it i.e.}
\begin{eqnarray}
\label{Lims}
\big[ \lim_{p^2\to 0}, \, \lim_{\epsilon\to 0} \big] \not= 0.
\end{eqnarray}
On the other hand, if the limits are simultaneous ones we deal with the uncertainty of $[0]^0$ which should be somehow resolved.

The second subtlety is related to the limit $p^2\to 0$ and the replacement of Eqn.~(\ref{Rep-1}).
Namely, in order to avoid the mentioned uncertainty, we have to implement the limit $p^2\to 0$ before the possible replacement.
In this case, the limit of $p^2\to 0$ is well-defined operation and we finally obtain that
\begin{eqnarray}
\label{V-nV-4}
&& \lim_{p^2\to 0} \mathcal{I}(p^2) = (c.c.) \int_{0}^\infty d\lambda \lambda^{1-D/2} =
\frac{1}{2- D/2} \Big\{
\lim_{\lambda\to\infty} \lambda^{2-D/2} - \lim_{\lambda\to 0 } \lambda^{2- D/2}
\Big\}
\nonumber\\
&&
\equiv \int \frac{(d^D k)}{(k^2)^2} =\mathcal{V}_2.
\end{eqnarray}

\section{$\delta(0)$-singularity}
\label{DeltaZ-1}

We are now in a position to discuss the treatment of $\delta(0)$-singularity (or $\delta(0)$-uncertainty).
To this aim,  we follow to the
sequential approach to the singular generated functions (distributions).

From one hand, based on the dimensional analysis, we may conclude that all massless vacuum integrations disappear, {\it i.e.}
\begin{eqnarray}
\label{V-in-1}
\mathcal{V}_n=\int \frac{(d^D k)}{[k^2]^n}=0 \quad \text{for}\,\, n\not= D/2.
\end{eqnarray}
However, the case of $n=D/2$ (or $n=2$ if $\varepsilon\to 0$) requires the special consideration because
the dimensional analysis argumentation does not now work. 
Nevertheless, the nullification of $\mathcal{V}_{D/2}$ takes still place but thanks to different reasons.
It turns out, the ultraviolet and infrared divergencies are cancelled each other.
Hence, if only the ultraviolet divergencies are under our consideration, $\mathcal{V}_{D/2}$ is not equal to zero.

To demonstrate, we dwell on the vacuum integration which is externally the IR-regularized one. 
It is necessary to remind that, in the space with $D=d-2\epsilon$, the positive value of $\epsilon$ allows to avoid the UV-divergency.
In the
spherical co-ordinate system, we write the following 
representation
\begin{eqnarray}
\label{V-in-2}
\mathcal{V}_{2}=\int_{UV} \frac{(d^D k)}{[k^2]^2}\equiv
 \frac{\pi^{D/2}}{\Gamma(D/2)} \int_{\mu^2}^{\infty} d\beta \beta^{D/2-3} \quad \text{with}\,\,\, \beta=|k|^2,
\end{eqnarray}
where $\mu^2$ plays a role of IR-regularization and the angular integration given by the measure $d\Omega$ is calculated explicitly.
Next, calculating $\beta$-integration, we reach the representation as
\begin{eqnarray}
\label{V-in-3}
\mathcal{V}_{2}=
 \frac{\pi^{2-\varepsilon} \mu^{-2\varepsilon} }{\Gamma(2-\varepsilon)}  \, \frac{1}{\varepsilon} \Big|_{\varepsilon\to 0},
\end{eqnarray}
where it is shown that the $\epsilon$-pole corresponds to the UV-divergency only because the IR-divergency is absent 
by construction thanks for $\mu^2$.
This is a very-well known representation used, for example, in \cite{Grozin:2005yg, Grozin:2007zz}.

On the other hand, we are able to calculate the vacuum integration by
Gorishni-Isaev's method \cite{Gorishnii:1984te}. In this case, $\mathcal{V}_n$ reads
\begin{eqnarray}
\label{V-in-4}
\mathcal{V}_{n}=\int \frac{(d^D k)}{[k^2]^n} =
\frac{2i\, \pi^{1+D/2}}{(-1)^{D/2}\, \Gamma(D/2)} \delta(n-D/2).
\end{eqnarray}
Supposing $D=4-2\varepsilon$, the only contribution is given by
\begin{eqnarray}
\label{V-in-4-2}
\mathcal{V}_{2}=\int \frac{(d^D k)}{[k^2]^2} =
\frac{2i\, \pi^{3-\varepsilon}}{ \Gamma(2-\varepsilon)} \delta(\varepsilon) \, \not=\, 0.
\end{eqnarray}
Hence, the delta-function of argument $\epsilon$ reflects the UV-divergency.
We specially stress that the representations of $\mathcal{V}_2$ given by Eqns.~(\ref{V-in-3})  and (\ref{V-in-4-2})
are equivalent.

The delta-function as a generated function (distribution) is a linear singular functional (which cannot be generated by any locally-integrated functions) defined on the suitable finite function space.
Such a definition is absolutely well but it is not unique one. Namely, the delta-function can be understood with the help of the fundamental sequences of regular functionals provided the corresponding weak limit, 
see for example  \cite{Antosik:1973, Gelfand:1964}.
Besides, one of the delta-function representations is related to the following realization
\begin{eqnarray}
\label{Delta-Real}
\delta(t)=\lim_{\varepsilon\to 0} \delta_\varepsilon(t)\equiv
\lim_{\varepsilon\to 0}  \frac{ St.F. (-\varepsilon \le t \le 0)}{\varepsilon},
\end{eqnarray}
where $St.F.(-\varepsilon \le t \le 0)$ implies the well-known step-function without any uncertainties.

Going back to Eqn.~(\ref{V-in-4-2}), one can see that the treatment of $\delta(\varepsilon)$ as the linear (singular)
functional on the finite function space with $d\mu(\varepsilon)=d\varepsilon \phi(\varepsilon)$ meets some difficulties
within the dimensional regularization approach. Indeed, for the practical use,
$\varepsilon$ is not a convenient variable for the construction of the finite function space because we finally need 
to be focused on the limit as $\epsilon\to 0$.

Meanwhile, within the sequential approach \cite{Antosik:1973, Gelfand:1964}, the delta-function might be considered as the usual singular (meromorphic)
function and  the $\delta(0)$-singularity/uncertainty can be treated as 
a pole of the first order \cite{Anikin:2020dlh},
\begin{eqnarray}
\label{D-treat}
\delta(0)=\lim_{\varepsilon\to 0}  \delta_{\varepsilon}(0)\equiv \lim_{\varepsilon\to 0}  \frac{1}{\varepsilon}.
\end{eqnarray}
For the demanding mathematician, the representation of Eqn.~(\ref{D-treat})  should be understood merely as a symbol.
That is, $\delta(0)$ denotes alternatively the limit of $1/\epsilon$.
This representation is also backed by the obvious fact that Eqns.~(\ref{V-in-3})  and (\ref{V-in-4-2})
are equivalent ones.

It is worth to notice that representation of $\delta(0)$ through the pole of an arbitrary meromorphic function
should be used very carefully.
For example, if we suppose that (here, $z\in \mathbb{E}^4$ and the delta-function is assumed to be a functional on 
the finite function space)
\begin{eqnarray}
\label{Ef-ex-1}
\big[ \delta(z)\big]^2 = \delta(0) \, \delta(z),
\end{eqnarray}
the representation given by 
\begin{eqnarray}
\label{Ef-ex-2}
\delta(z) = \lim_{\epsilon\to 0} \delta_\epsilon(z), \quad 
\delta_\epsilon(z) = \frac{1}{\pi^2 \epsilon^4}
e^{- z^2/\epsilon^2} \Rightarrow \delta(0) \sim \delta_\epsilon(0)=\frac{1}{\pi^2 \epsilon^4}
\end{eqnarray}
does not satisfy the condition of Eqn.~(\ref{Ef-ex-1}).
Another informative example can be found in \cite{Efimov:1973pjo}.

\section{Conclusion}
\label{Sec:Con}

To conclude, we have presented the important explanations regarding the massless vacuum integrations. 
In the note, we have demonstrated the preponderance of sequential approach where 
the singular generated functions (distributions) are treated as a fundamental  
sequences of regular functionals. Due to this treatment, the uncertainty as $\delta(0)$ can be resolved
via the meromorphic function of first order.
Also, it has been shown in detail how the delta-function represents either UV-regime or IR-regime.

\section*{Acknowledgements}

Our special thanks go to S.V.~Mikhailov and L.~Szymanowski for very useful and stimulating discussions.


\end{document}